\documentstyle[11pt,aaspp4,flushrt]{article}

\slugcomment{To appear in the Astronomical Journal}

\lefthead{}
\righthead{}

\begin{document}

\title{Evolution of the Galaxy Population Based on Photometric Redshifts in the 
Hubble Deep Field}

\author{M. J. Sawicki, H. Lin, and H. K. C. Yee}
\affil{Department of Astronomy, University of Toronto, Toronto, 
Ontario M5S 3H8, Canada
\\ email: sawicki, lin, hyee @astro.utoronto.ca}


\begin{abstract}

This paper presents the results of a photometric redshift study of
galaxies in the Hubble Deep Field (HDF).  The method of determining
redshifts from broadband colors is described, and the dangers inherent
in using it to estimate redshifts, particularly at very high $z$, are
discussed.  In particular, the need for accurate high-$z$ spectral
energy distributions is illustrated.  The validity of our photometric
redshift technique is demonstrated both by direct verification with
available HDF spectroscopic data and by comparisons of luminosity
functions and luminosity densities with those obtained from $z<1$
spectroscopic redshift surveys.  Evolution of the galaxy population is
studied over $0\lesssim z < 4$.  Brightening is seen in both the
luminosity function and the luminosity density out to $z\approx 3$;
this is followed by a decline in both at $z>3$.  A population of
$z<0.5$ star-forming dwarfs is observed to M$_{\rm F450W_{AB}}=-11$.
Our results are discussed in the context of recent developments in the
understanding of galaxy evolution.

\end{abstract}




%

\section{INTRODUCTION}


The study of galaxy evolution is a fundamental but difficult
problem in cosmology. 
Only recently have relatively large surveys, even at modest and intermediate
redshifts ($0.2 \lesssim z \lesssim 1.5$),
been carried out (e.g., \cite{lil95}; \cite{cow96};
\cite{ell96}; \cite{gla95b}; \cite{yee96b}).
Lilly et al.\ (1995) and Ellis et al.\ (1996) have studied the galaxy
luminosity function (LF) and shown that it evolves over $0 \lesssim z
\lesssim 1$.  Lilly et al.\ report that the LF for red galaxies does
not evolve significantly over $0 \lesssim z \lesssim 1$, and conclude that a
population of red massive objects must have already been ``in place''
by $z\approx 1$.  On the other hand, their blue subsample is evolving,
which indicates that its member galaxies formed later than those in
the red subsample.  Out to somewhat higher redshifts ($z<1.6$) Cowie
et al.\ (1996) have observed a population of actively star-forming,
massive galaxies, and have also found evidence for
``downsizing''--- a trend in which more massive galaxies appear to be
forming at higher redshifts.  Morphological studies (\cite{sch95};
\cite{sch96}) show that disks brighten with increasing redshift,
indicating a higher rate of star formation in the past.  Quite
recently, a number of large, star-forming galaxies have been discovered at
high redshifts ($z \approx 3$), either serendipitously
(e.g. \cite{yee96}), or as part of targeted searches
(e.g. \cite{ste96}).  These high-redshift objects could very well be
the progenitors of present-day typical massive galaxies.  These and
other clues (see \cite{fuk96} for a recent review) hint at a scenario
in which massive galaxies formed at $z\approx 3$, and were then
followed by a sequence of less and less massive galaxies forming at
lower and lower redshifts, leading down to the formation of dwarfs at
recent ($z<0.5$) epochs.

To obtain a more definitive and complete picture of the star formation
history of galaxies requires systematic surveys covering the full redshift 
range from the epoch of galaxy formation to the present.
At higher redshifts such an
investigation is difficult and time-consuming, as galaxies are faint
and spectroscopic features difficult to observe. Until recently, known
high-redshift galaxies were atypical objects characterized by unusual
activity such as strong radio emission (\cite{mca93}).  More indirectly, 
one often relies on Lyman and
metal-line absorption systems which are assumed to be the progenitors
of present-day disks and spheroids (e.g. \cite{lan95}).  The current
numbers of relatively normal, spectroscopically observed galaxies at
high redshifts ($z \approx 3$) is still small (e.g.,
\cite{ste96}). However, 
the
combination of very deep multi-color images and the photometric
redshift technique can provide a relatively easier and less
time-consuming way of exploring the high-redshift universe and the
evolution of galaxies.

The Hubble Deep Field (HDF)\footnote{Based on observations with the
NASA/ESA \emph{Hubble Space Telescope} obtained at the Space Telescope
Science Institute, which is operated by the Association of
Universities for Research in Astronomy, Inc., under NASA contract NAS
5-26555.}  is a very deep set of four-band imaging exposures of a
``random'' high galactic latitude field (\cite{wil96}).  The images
were obtained by the Hubble Space Telescope (HST) at the end of 1995
and were made available to the community shortly thereafter.  Because
of its extreme depth (limiting F814W$_{\rm AB}$ magnitude $>28$), wide
wavelength coverage (3000--8000\AA), and excellent spatial resolution,
the HDF affords an unprecedented opportunity to study the galaxy
population to unmatched lookback times.

The redshifts of objects seen in the HDF are of great importance to
studies of galaxy evolution, and extensive programs of spectroscopic
observations have been undertaken in order to secure them (e.g. \cite{coh96};
\cite{mou96}).
Unfortunately, because of the HDF's extreme depth, spectroscopic
redshifts are not practical for all but the brightest objects in that
field.  One can, however, use the colors of galaxies to estimate their
redshifts with a fair degree of confidence.  This color-, or
photometric-, redshift technique (e.g. \cite{loh86}; \cite{con95})
allows the determination of redshifts for HDF objects too faint to be
spectroscopically accessible.  A number of authors have recently
applied the photometric redshift technique to the Hubble Deep Field.
Lanzetta, Yahil, \& Fern\'andez-Soto (1996) have used color redshifts to
identify protogalaxy candidates at $z>5$.  At slightly lower
redshifts, Gwyn \& Hartwick (1996) ($z<5$) and Mobasher et al.\ (1996)
($z<3$) have used their photometric redshifts to generate luminosity
functions.  In this paper we present our photometric redshift
measurements and then study the luminosity function and luminosity
density evolution to $z=4$.  Our results differ markedly from those in
the Gwyn \& Hartwick and the Mobasher et al.\ studies.

In Section~\ref{data} we briefly describe the data and the photometric 
measurements.
We then discuss the determination of photometric redshifts  
(Section~\ref{technique}) and the possible pitfalls along the way 
(Section~\ref{aliasing}).  
Using what we consider to be our most reliable photometric redshifts,
we go on to compute the luminosity function (Section~\ref{lumfuncs})
and luminosity density (Section~\ref{lumdens}), and discuss the
observed evolution in the context of a specific current picture of 
galaxy formation (Section~\ref{discussion}).
Throughout this paper we assume a flat, matter dominated universe
($q_0 = 0.5,\ \Omega_0 = 1$) with $H_0 = 100 \ h \ {\rm km \ s^{-1}
\ Mpc^{-1}}$. We use $h = 1$ if not otherwise indicated.


\section{DATA}\label{data}

The HDF has been observed in four broadband filters (F814W, F606W,
F450W, and F300W; central wavelengths of 8140, 6060, 4500, and
3000\AA\ respectively).
We used publicly available Version 2 images of the HDF.
These images have been processed using the drizzling technique 
and have a pixel size of 0.04 arcsec (\cite{wil96}).
We chose to work in the AB system\footnote{An approximate conversion
between the AB and Vega-based systems is F300W$_{\rm AB}=$ F300W$_{\rm
Vega}+1.33$, F450W$_{\rm AB}=$ F450W$_{\rm Vega}-0.08$, F606W$_{\rm
AB}=$ F606W$_{\rm Vega}+0.12$, and F814W$_{\rm AB}=$ F814W$_{\rm
Vega}+0.44$} (\cite{oke83}), using the STScI zero-points given by
Ferguson (1996).  
We used the three Wide Field Camera images which we trimmed because
image quality degrades significantly near the edges; all told, the
angular area used was 4.48 arcmin$^2$.

We performed object finding and photometry using the PPP faint galaxy
photometry package (\cite{yee91}).  Automatic object finding was done
in both the F814W and F606W frames, with the results edited by
eye and then combined into a single catalog; 1620 objects were thus
detected to our F814W$_{\rm AB}$ completeness limit of $28.0$.  
Of these 1620, 43 were morphologically classified as stars and
were not included in any subsequent analysis.  
At faint apparent magnitudes many objects are detected in fewer than
all four HDF filters; the fraction of objects detected in all four
bands is 500/529, 891/1003, and 1289/1577 for objects brighter than
F814W$_{\rm AB}=$ 26, 27, and 28, respectively.  
We confined our analysis to objects with F814W$_{\rm AB} \leq 27$.

Because accurate determination of galaxy colors is essential if one is
to use them to determine redshifts, we briefly outline the way in
which photometry is done in PPP.  PPP analyzes the flux growth curves
and determines an ``optimal aperture'' for each filter.  Since, of the
four HDF bands, the F606W images are the deepest, the F606W flux
within the optimal aperture is used to derive the fiducial ``total
magnitude'' of the object.  The object's color is determined using a
``color aperture'', which is the smallest of (1) the optimal aperture
for that given filter, (2) the F606W optimal aperture, and (3) an
aperture of 1.56$"$ (39 pixels) diameter.  The above procedure ensures
that the measurement of the spectral energy distribution of an object
is done over an identical angular area in the four bands.  The color
aperture (which is generally smaller than the optimal aperture) is
used to improve the signal-to-noise ratio in the measurement of the
object's color.  ``Total magnitudes'' for the other three filters are
then obtained by correcting the F606W magnitude with the measured
colors.  Strictly speaking, this procedure will produce the
``correct'' total magnitudes for the other filters only if there is no
color gradient in the galaxy.  However, we note that the uncertainty
introduced by this procedure is small for most galaxies relative to
the total photometric uncertainty.  Typical photometric uncertainties
at F814W$_{\rm AB}=27$ are 0.10, 0.09, 0.17, and 0.45 in F814W$_{\rm
AB}$, F606W$_{\rm AB}$, F450W$_{\rm AB}$, and F300W$_{\rm AB}$
respectively.


Ironically, object finding in the HDF may suffer from too large a
lookback time and too high a resolution.  It is a matter of contention
whether a small object in the vicinity of a large galaxy is classified as 
an individual faint galaxy, or as a fragment or HII region belonging to the 
nearby parent (see \cite{col96}).  Since the presence of substructures
misclassified as faint galaxies could affect our results, we inspected the 
environments of faint ($26< {\rm F814W_{AB}} <27$) objects.  
It was found that, in our catalog, $\sim
70$\% of these objects have absolutely no large companions.  Of
the remaining $\sim 30$\%, most appear to be associated with larger 
objects only in projection.  Even if as many as $\sim 30$\% of 
faint objects are indeed components of larger galaxies, the decrease 
in the faint galaxy numbers is
insufficient to significantly affect our results.

\section{DETERMINATION OF REDSHIFTS}\label{technique}

\subsection{Color Redshifts}

The redshift of a galaxy can be estimated by comparing its observed
broadband spectral energy distribution (SED) with a set of template
SEDs for galaxies at different redshifts and of different spectral types.  
The technique, termed ``photometric'' or ``color redshifts,''
can be thought of as very-low-resolution spectroscopy.  Various 
implementations of the
technique have been applied to both cluster and field galaxies at $z<1$
(e.g. \cite{loh86};
\cite{con95}; \cite{bel95}).

As Connolly et al.\ (1995) point out, the spectral feature that is
most important for determining the photometric redshift of a galaxy at
$z<1$ is the 4000\AA\ break.  The size of the break and the curvature
of the spectrum to either side of it carry information about the
galaxy's spectral type.  Because of the HDF's expected redshift depth
($z>1$) and UV coverage, spectral breaks other than those seen at
rest-frame optical wavelengths will be important.  In particular, the
912\AA\ Lyman break will redshift into the F300W filter at $z\approx
2.3$.  Features at $\sim$3646\AA\ (the Balmer break), $\sim$2800\AA\
(due to Mg II), and $\sim$2635\AA\ (due to Fe) will also play a role,
particularly in older stellar populations.  In Section
\ref{aliasing} we examine the effects that misidentification of the
various spectral breaks may have on the redshifts measured and the 
conclusions subsequently reached regarding galaxy evolution.

\subsection{Measurement of Redshifts and Spectral Types}

We determine the best-fitting redshift and spectral type for each HDF
object by comparing its colors against those of a set of templates
spanning a range of redshifts (0--5) and spectral types (star-forming
through old stellar population).  The construction of template colors
for the HDF is hampered by the fact that at $z>1$ one observes
rest-frame UV regions of the objects of interest---regions for which
spectral energy distributions are poorly known even for local
galaxies, but which may suffer strong internal
reddening.  A second complication is that at high $z$ the UV flux is further 
suppressed through absorption by intergalactic gas (e.g. \cite{mad95}).

\subsubsection{Technique}

We constructed templates in the following way: We used the SEDs of
Coleman, Wu,~\& Weedman (1980; hereafter CWW) which we augmented with
two very blue SEDs.  The CWW SEDs are a collection of empirical SEDs
of representative local galaxies ranging from E to Im in spectral
type, and covering wavelengths from 1500 to 10000\AA.  We extended these
SEDs below 1500\AA\ as follows: For spectral types E and Sbc the
extension is a power-law extrapolation of the 1500--2500\AA\ region of
the SED.  For spectral types Scd and Im we extrapolated a power law as 
for the earlier spectral types, but then replaced it with a GISSEL model 
spectral shape which has been normalized to that power law extrapolation; 
the aim of this replacement was simply to reproduce the 912\AA\ break 
which should be quite
prominent in the later spectral types, but which would not be reproduced in
a simple power-law extrapolation.
The two very blue SEDs with which we augment the CWW
set were generated from the GISSEL library and represent young,
star-forming galaxies; specifically, we used the constant star
formation rate models with a Salpeter IMF (masses $0.1 M_{\sun} < M < 125
M_{\sun}$) and ages of 0.5 and 0.05 Gyr.
These two very blue SEDs were added because there are, even locally,
substantial numbers of galaxies which are bluer than the bluest CWW
type.  

We then interpolated between our extended CWW SEDs to cover the
spectral-type range more finely.  To predict the observed SEDs for a
given redshift, we applied the Lyman continuum and line blanketing
suppression of the UV flux due to intervening Lyman-forest and
Lyman-limit absorbers (\cite{mad95}).  Finally, template colors were
constructed by convolving our modified CWW SEDs with the HST
instrumental response curves.  At the end of this process we have a
set of templates as a function of redshift and spectral type.
We will from now on refer to this template set as the ``extended CWW''
set.  It is this template set which was used to generate the redshifts
utilized in the analyses of Sections \ref{lumfuncs} and \ref{lumdens}.

The extended CWW template set described above assumes that the
spectrum of a galaxy of a given spectral type does not evolve with
time.  This assumption is obviously questionable considering that at
redshifts of interest the universe is only a fraction of its present
age.  A straightfoward way to treat the problem of the evolving SED
would have been to use, as Gwyn \& Hartwick (1996) did, spectral
evolution models such as the GISSEL library.  However, spectral
synthesis models do not account for a galaxy's internal absorption
which is particularly important in the UV (see Section
\ref{aliasing}).  Furthermore, there are considerable differences amongst 
the SEDs predicted by differrent spectral synthesis models (see
\cite{cha96}).  For these reasons we chose not to use spectral synthesis 
models as templates, preferring instead the non-evolving, empirical 
CWW spectra. 


We note, however, that early in galaxy formation the SEDs of all 
spectral types are dominated by massive stars and consequently are
well represented by late spectral type CWW SEDs.  As a galaxy ages due
to the aging of its stellar population, its SED will migrate across
the range of CWW SEDs from later to earlier spectral types.  If the
galaxy is actively star-forming, then its SED will of course remain
similar to a late-type CWW SED.  
Consequently, our extended CWW templates 
can be thought of as corresponding to stellar populations 
of different ages rather than to different galaxy morphological types. 
We therefore believe that our
extended CWW SEDs are a fair representation of real-life evolving SEDs.  

We used $\chi^2$ fitting to find the best-matching redshift and
spectral type for each HDF object.  For each template we calculated
\begin{equation}
\chi^2 =\sum_{i}\left[{ \frac{F_{observed,i}-s\cdot F_{template,i}}{\sigma_{i}}}\right]^2.
\end{equation}
Here, $F_{observed,i}$ is the flux observed in a given filter $i$ and
$\sigma_{i}$ is its uncertainty; $F_{template,i}$ is the flux of the
template in the same filter.  The scaling term $s$ normalizes the
template to the observed SED, and the sum is taken over all four HDF
filters.  The best-matching redshift and spectral type are obtained by
minimizing $\chi^2$ as a function of template and $s$.  
Recall that some objects (112/1003)
were detected in fewer than all four filters.  Such objects were fitted 
using the upper bound on the object's flux:  it was assumed that 
$F_{observed}$ and $\sigma_{observed}$ were both equal to the
1$\sigma$ detection limit for that object and filter and the fit was done 
as before.

\subsubsection{Results}

In Figure~\ref{fig_zphot_vs_ztrue} we compare our photometric
redshifts against 74 HDF spectroscopic redshifts available from various
sources (\cite{hog96}; \cite{ste96}; \cite{phi96}).
The agreement is good with a scatter of $\sigma_z =
0.12$ for the $z<1.5$ objects, increasing to $\sigma_z=0.28$ for those
at $z>2$.  The catastrophic failure rate is small, with only 2 out of
55 $z<1.5$ objects being assigned anomalously high photometric
redshifts. 

\begin{figure}
\plotone{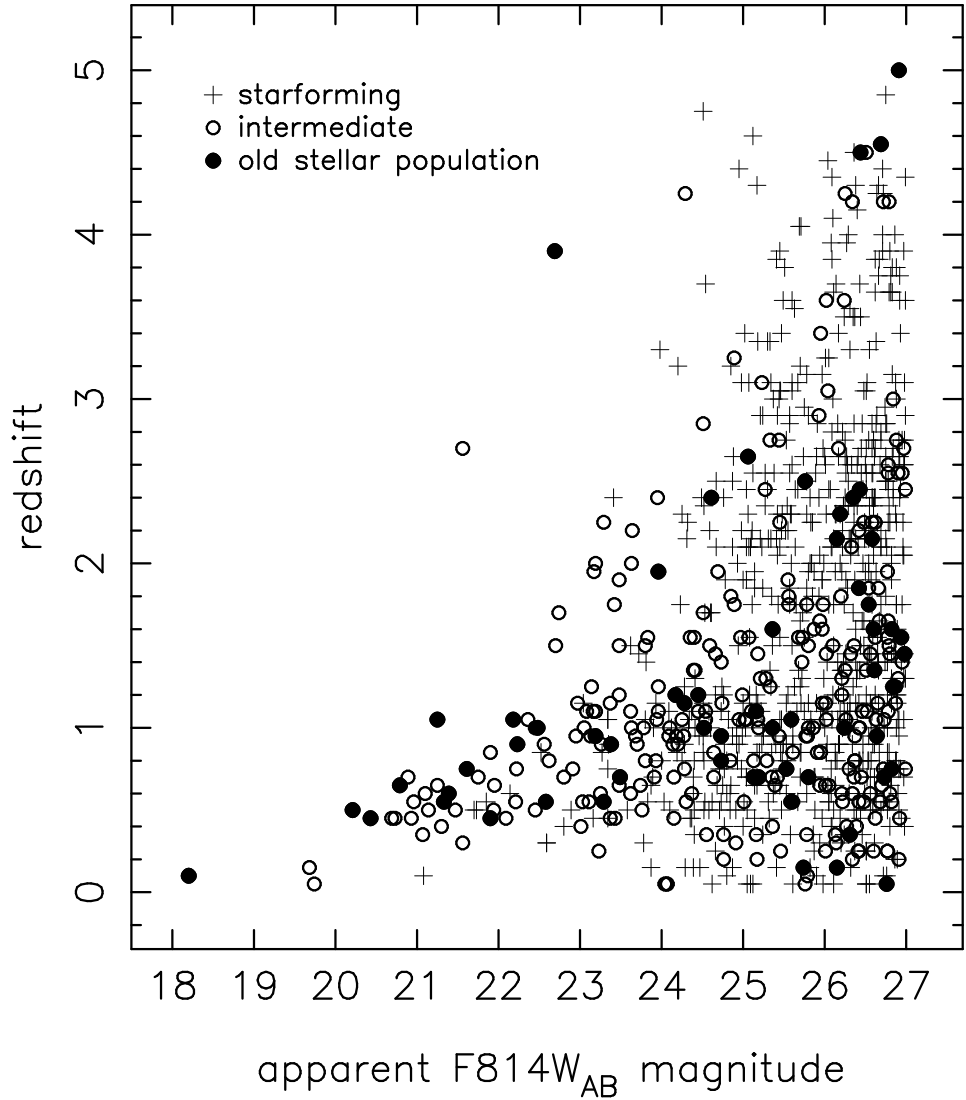}
\caption{
The Hubble Diagram for the HDF.
 Different symbols correspond to ranges of spectral types: filled
 circles are galaxies with an old stellar population, open circles
 are objects with intermediate ages, and crosses denote
 star-forming galaxies.}
\label{fig_hubble_diagram} 
\end{figure}

Figure~\ref{fig_hubble_diagram} shows the Hubble diagram for the HDF.
The different symbols refer to different spectral types: filled
circles denote spectral types with an old stellar population (CWW E
through other early SEDs),
open circles are objects with
stellar populations of intermediate age (CWW Sbc through Scd)
while crosses are star-forming objects (CWW Im 
through the two very blue SEDs).
We draw attention to the presence, at $4<z<5$, of a small number of objects
with early spectral types.  We regard these objects as low-redshift
galaxies whose redshifts have been aliased (see
Section~\ref{aliasing}) to high redshifts because of the confusion
between the 4000\AA\ and 912\AA\ breaks.

Figure~\ref{fig_zdist} presents the redshift distribution of the HDF
galaxies (thick line) for two apparent limiting magnitudes.  The
sample is also split by spectral type.  Note that there is a prominent
peak at $z\lesssim 1$ which is strongly dominated by star-forming
galaxies (dashed line) at faint magnitudes (top panel).  A second
population of star-forming galaxies can be seen at $z\approx 2$--$2.5$
in the top panel of Figure~\ref{fig_zdist}.  Though there is a slight
increase in the overall number of galaxies at $z\approx 2.2$, the very
large high-$z$ peak reported by Gwyn \& Hartwick (1996) is not seen in our
analysis.

\section{REDSHIFT ERRORS}\label{aliasing}

The photometric redshift technique will determine an 
incorrect redshift when the colors of an observed galaxy match the
colors of a ``wrong'' template more closely than they match those of
the ``right'' one.  Such aliasing may happen for three reasons: (1)
random photometric errors, (2) a template set that is too sparse, and
(3) a template set produced with unrealistic SEDs.

\begin{figure}
\plotone{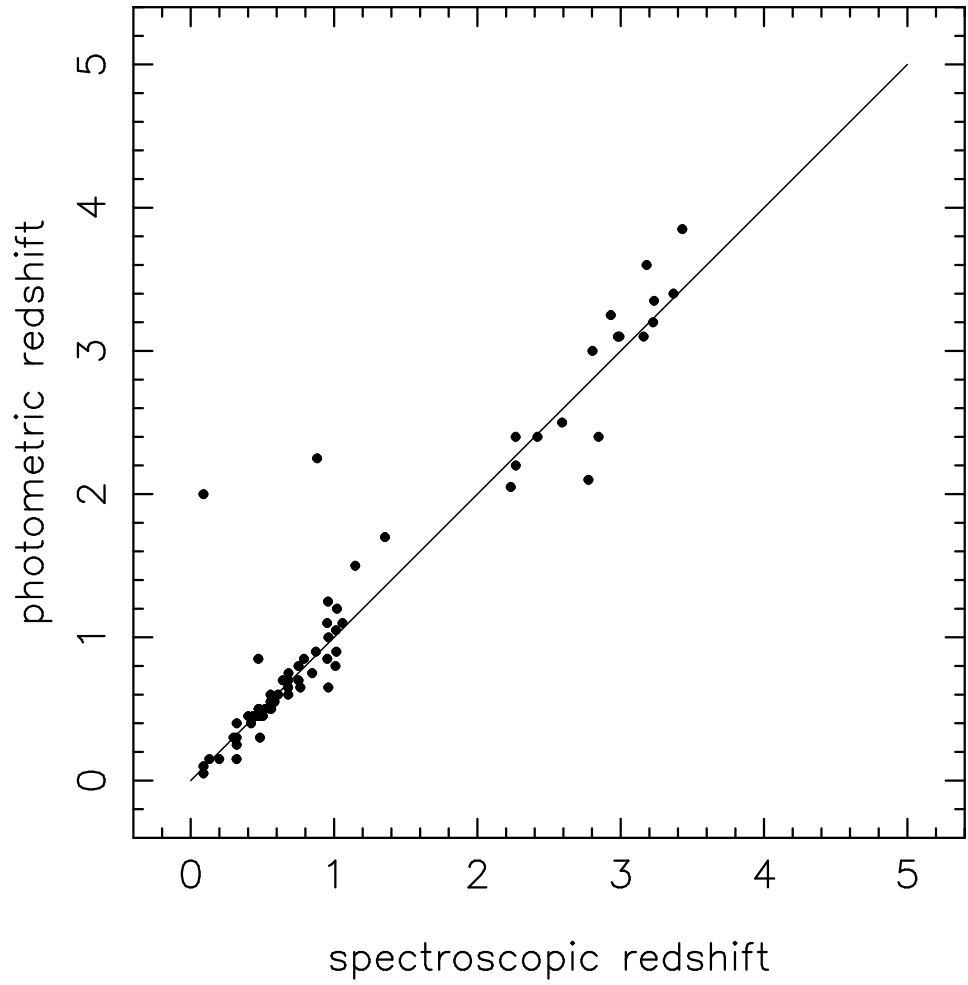}
\caption{Comparison of photometric and
spectroscopic redshifts.  Photometric redshifts were obtained using
our extended CWW template set.  A perfect match would lie on the
 diagonal line.}
\label{fig_zphot_vs_ztrue}
\end{figure}

Aliasing of templates produces errors which are either minor or
dramatically catastrophic.  Minor redshift errors (such as most of the
redshift discrepancies seen in Figure~\ref{fig_zphot_vs_ztrue}) are
best described as ``noise'' and are relatively benign.  The size of
this noise is on the order of $\sigma_z=0.1$--$0.3$ as can be seen in
Figure~\ref{fig_zphot_vs_ztrue}.  If need be, the effect of the
redshift noise can be modeled (as is done in Section~\ref{lumfuncs})
by means of Monte Carlo simulations.

Catastrophic errors can, however, have a profound effect on the
conclusions one reaches regarding galaxy evolution (as will be
illustrated in Section~\ref{wrong_templates}) and so we turn to
investigate their nature more closely.

\subsection{Effects of Random Photometric Errors}

If the colors of an observed galaxy are similar to the colors of 
\emph{two} templates,  then random photometric errors may
tip the scales in favor of one or the other of the templates.  If
the two templates have vastly different redshifts, one of them right
and one wrong, then the observed object may be assigned a
catastrophically erroneous redshift.  

One might expect the \emph{fraction} of objects scattered from low
redshift to high redshift to be the same as that scattered from high
redshift to low redshift.  However, the \emph{number} of objects
scattered from low redshift to high redshift is likely to be much
larger than that scattered in the opposite direction since one can
expect there to be many more objects (down to a certain apparent
magnitude) at low redshift than at high redshift.

The importance of this type of redshift aliasing depends on the size
of photometric uncertainties.  Monte Carlo simulations show that
catastrophic aliasing of redshifts in the HDF is insignificant for
bright objects and starts to become noticeable only at
F814W$_{\rm AB} \approx 27$ at the 10\% level.  It is for this reason that 
we limit our analysis to objects brighter than F814W$_{\rm AB} = 27$.

\subsection{Effects of a Sparse Template Set}

Aliasing of redshifts may also occur when the template set is sparse.
If templates at the correct redshift are placed too far apart (i.e.\ too
sparsely) in spectral type, then it may happen that a template with a
catastrophically wrong redshift matches the observed SED more
closely than does either of the two best-matching templates at the
correct redshift.  In such a case the wrong redshift (and possibly
spectral type) will be chosen.
In addition to the spectral type dimension, the template set can also
be too sparse in the redshift dimension.  Likewise, the same
sparseness problem may occur if the template normalization term ($s$ in
equation(1)) with too coarse a step size is used in the fitting. 

The sparseness aliasing problem can be easily avoided by using
sufficiently many intermediate spectral types, redshifts, and steps in
$s$.  In particular, we used 81 spectral types, redshift steps of
0.05, and $s$ steps of 0.1 magnitude.  Decreasing the sparseness of
the template set further had no effect on the redshifts and spectral
types that were measured.

\subsection{Effects of an Unrealistic Template Set}\label{wrong_templates}

The photometric redshift technique is likely to identify the correct
redshift and spectral type provided the grid of templates includes a
template which is at the correct redshift \emph{and} matches the
observed colors of the galaxy.  If, however, the ``correct'' template
is not included in the template set, an erroneous redshift and
spectral type may be chosen.  As Connolly et al.\ (1995) point out,
the photometric redshift ``signal'' (at low redshift) comes from the
4000\AA\ break.  As we noted earlier, other spectral breaks exist at
3656\AA, $\sim$2635\AA, $\sim$2800\AA, and 912\AA.  Since strong
spectral breaks are primary sources of signal for identifying
photometric redshifts, one can expect that there will be ambiguity
between the various breaks and, hence, the corresponding redshifts.
One then has to rely on the relative break sizes and on spectral
curvature to identify the correct redshift.  If the template set being
used is erroneous in the sense that the break sizes and spectral
curvatures do not model those in real galaxies, one may well be
systematically identifying incorrect redshifts.

\subsubsection{The Cause}

UV spectra are not well known even for local galaxies.  Since it
is in the rest-UV that one observes extremely high redshift galaxies,
one has to be concerned about the effect that this will have on the
correctness of one's template set.

The problem is compounded further because galaxies evolve on
timescales comparable to the look-back times that one hopes to probe
in the HDF.  One could hope to include the effects of spectral
evolution by using spectral synthesis models such as the GISSEL
library (\cite{bru93}).  Such models, however, simulate the naked
stellar populations and do not include the effects of galactic
self-absorption which may strongly depress the flux blueward of
$\sim$1500\AA.  Furthermore, there is still considerable disagreement
between the various spectral synthesis models (see \cite{cha96}).
This disagreement is serious enough that templates constructed using
different models may give catastrophically different photometric
redshifts.

It is for these reasons --- uncertainty in evolutionary SED codes and 
their lack of internal reddening --- that we have chosen to base our
analysis on photometric redshifts obtained using templates that are 
based on local galaxy spectra (i.e. the ``extended CWW'' set). 

Yet another factor which can contribute to making one's template set
unrepresentative is intergalactic hydrogen.  This hydrogen, which
resides in Lyman-$\alpha$ clouds, will suppress UV flux through
continuum absorption and line blanketing (\cite{mad95}).  If one
neglects to account for this absorption, one will end up with an
unrealistic template set even if the input SEDs otherwise match those
of real high-redshift galaxies.

\subsubsection{The Effect}

To illustrate the effect that our poor knowledge of UV SEDs may have
on the values of photometric redshifts, we compare the redshifts
obtained using our ``best model'' --- the extended CWW --- with those
obtained using a template set based on pure GISSEL models.  We took the
($\Omega=1$) reference models of Pozzetti, Bruzual, \& Zamorani (1996)
which are GISSEL SEDs chosen to match local E/S0, Sab-Sbc,
Scd-Sdm, and ``very blue'' (rapidly star-forming) objects.  In every
respect (other than Lyman absorption which we did not apply) we have
processed the pure GISSEL SEDs of Pozzetti, Bruzual, \& Zamorani
(1996) in exactly the same way as in the case of the extended CWW
SEDs.

The pure GISSEL template set makes no correction for internal
absorption, nor does it account for high-$z$ Lyman absorption.  Even
though Gwyn \& Hartwick (1996) used evolving SEDs they noted that
using present-day SEDs made no difference to their results.  We
therefore assume that our pure GISSEL template set is similar to the
template set used by Gwyn \& Hartwick (1996).  Photometric redshifts
in the HDF were then computed with the pure GISSEL template set, in
the same fashion (same $z$ and $s$ step sizes) as those that were
obtained with the extended CWW template set.

\begin{figure}
\plotone{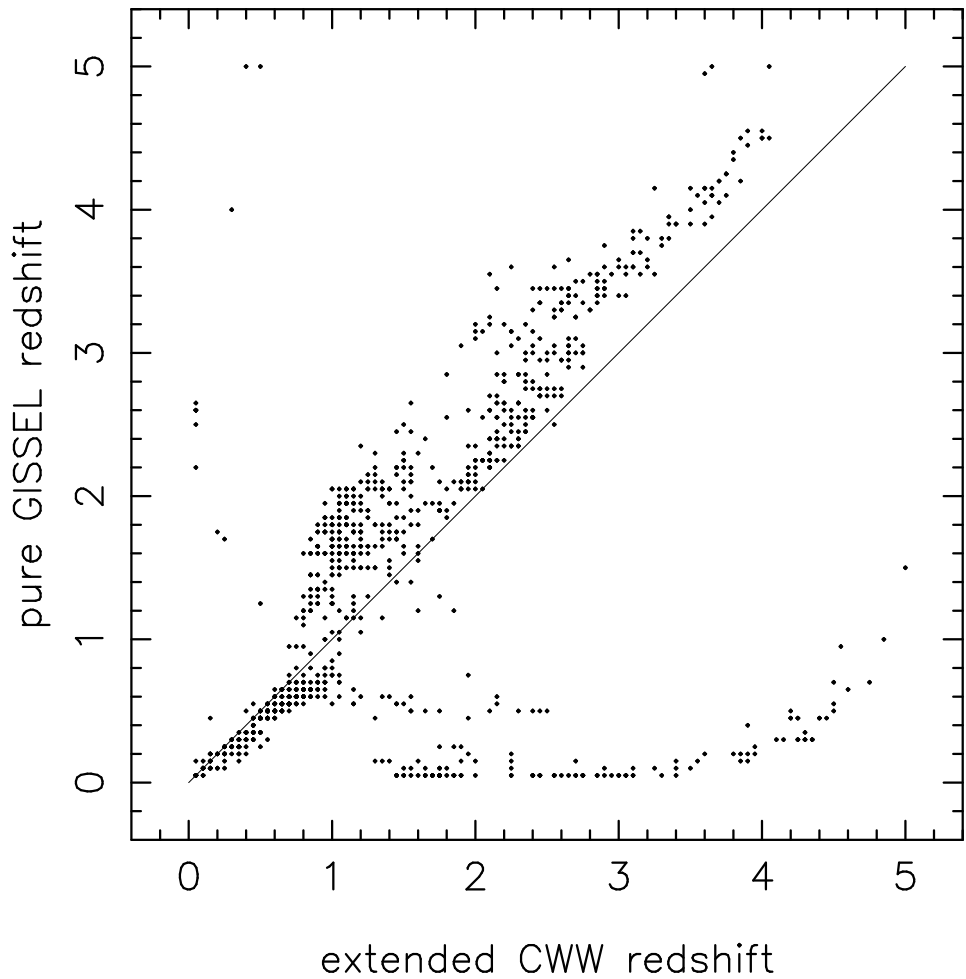}
\caption{
Comparison of photometric redshifts
 obtained using a pure GISSEL template set with those obtained using our
 ``best model'' extended CWW templates.  The pure GISSEL template set
 accounts for neither internal reddening nor high-$z$ Lyman
 absorption, both of which suppress the UV flux.  }
\label{fig_zCWW_vs_zGISSEL}
\end{figure}

Figure~\ref{fig_zCWW_vs_zGISSEL} compares the photometric redshifts
obtained using the pure GISSEL templates described above with those
obtained using our extended CWW set.  The two template sets produce
similar results up to $z\approx 0.8$ but then diverge considerably.
This divergence can be attributed to aliasing between spectral breaks
--- the 4000\AA\ break (giving a lower redshift in the extended CWW
fits) and a combination of Balmer, $\sim$2800\AA, and $\sim$2635\AA\
breaks (giving the higher redshifts in the pure GISSEL fits).  Overall,
only 55\% of redshifts obtained using the two template sets are within
0.5 of each other.  We believe that the extended CWW template set
produces more accurate redshifts, since, in contrast to the pure
GISSEL set, it accounts for the presence of reddening and Lyman
absorption.

\begin{figure}
\plotone{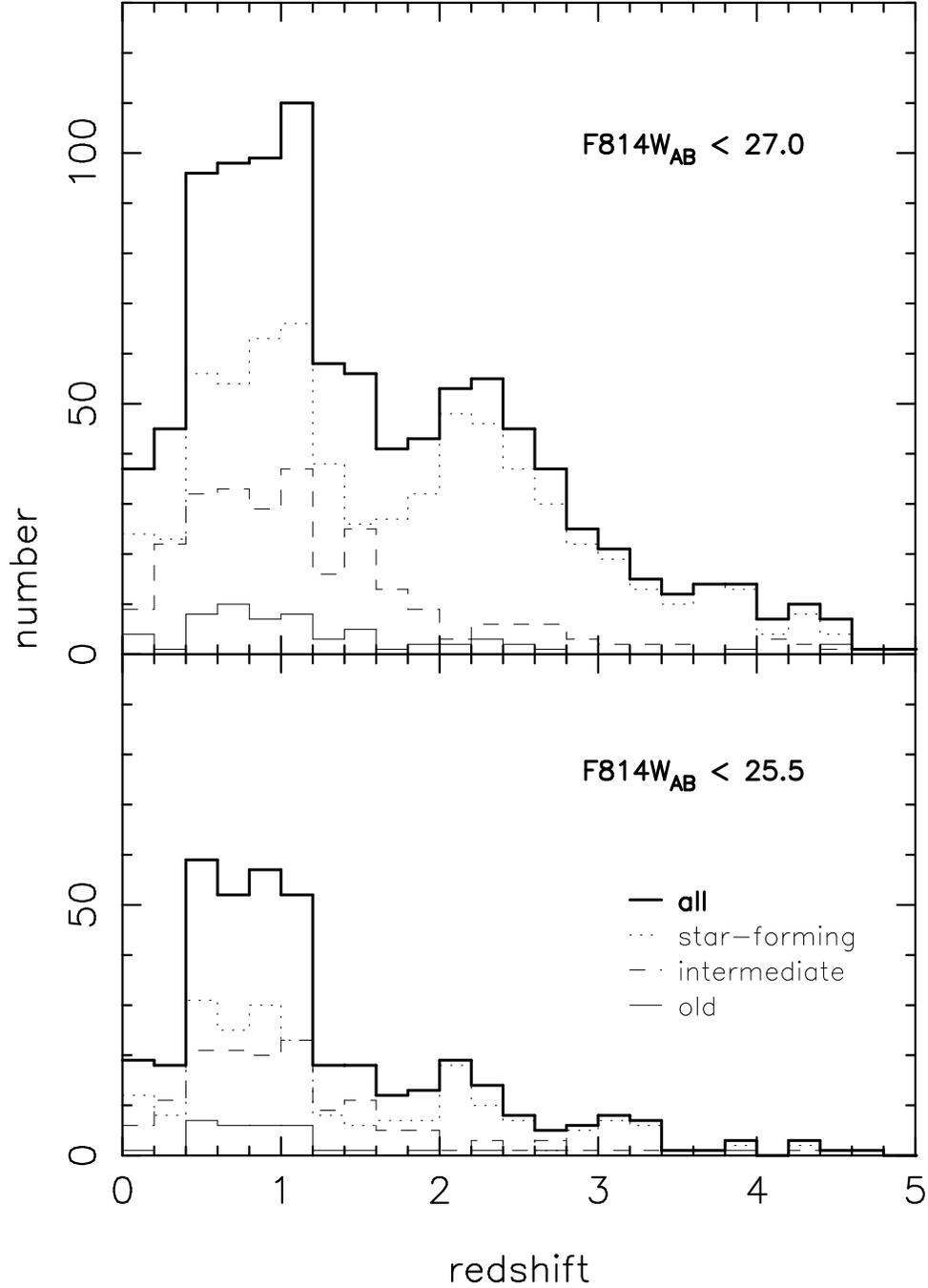}
\caption{
 Redshift distribution of objects brighter than
 F814W$_{\rm AB}=27$ (top panel) and those brighter than F814W$_{\rm
 AB}=25.5$ (bottom panel).  The thick solid line denotes all objects
 irrespective of spectral type.  The dotted line denotes star-forming
 galaxies, the dashed line denotes objects with intermediate ages, and
 the thin solid line denotes those with an old stellar population.}
\label{fig_zdist}
\end{figure}

\begin{figure}
\plotone{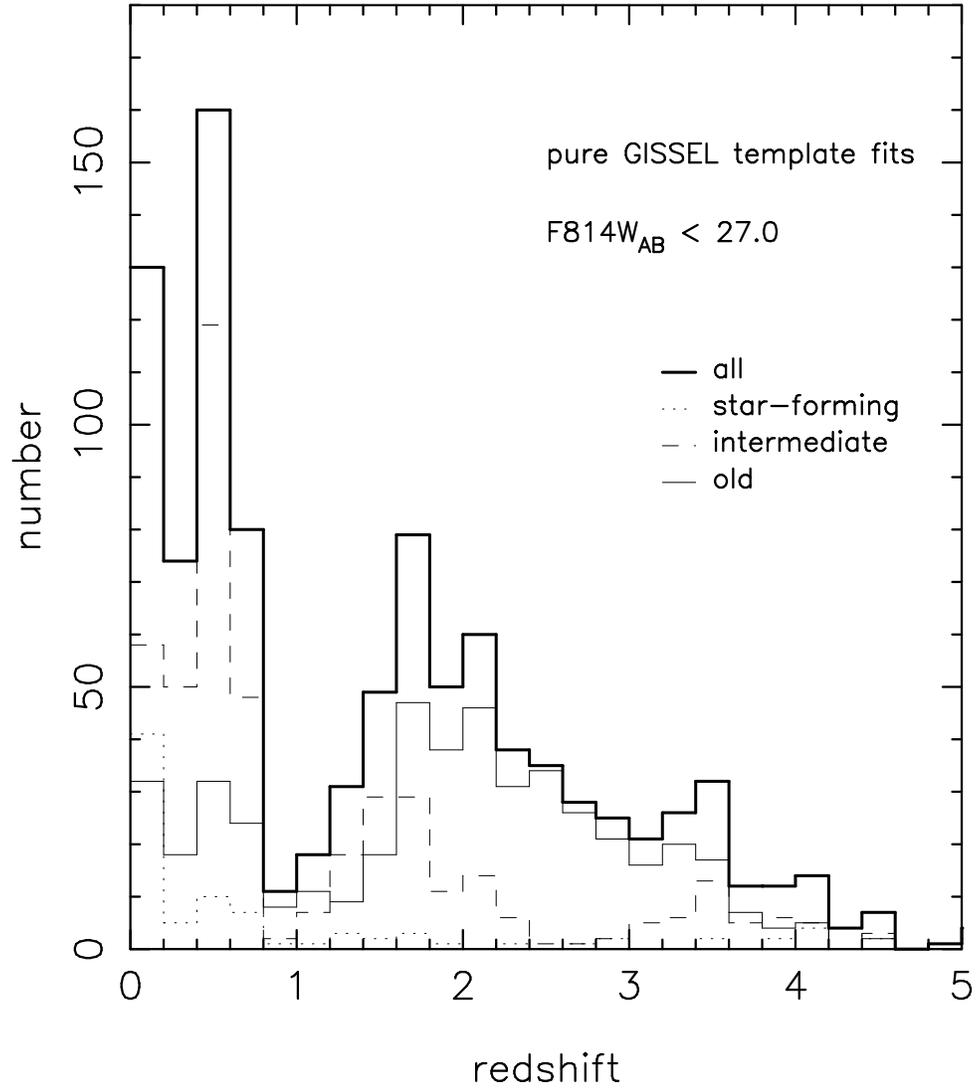}
\caption{
Redshift distribution (cumulative and
 split by spectral type) obtained using pure GISSEL templates which
 account for neither internal reddening nor high-$z$ Lyman
 absorption.  Symbols are the same as those in
 Figure~\ref{fig_zdist}.}
\label{fig_zdist_GISSEL}
\end{figure}

In addition to drastic differences in redshift, the two different
template sets assign completely different spectral types.  In
Figure~\ref{fig_zdist_GISSEL} we present the redshift distribution
obtained using the pure GISSEL templates.  This redshift distribution is
characterized by two prominent peaks, in contrast to the redshift
distribution shown in Figure~\ref{fig_zdist}.  The redshift
distribution subdivided by spectral type is also shown, using the same
convention as in Figure~\ref{fig_zdist}.  Note that the majority of
$z>1$ objects in Figure~\ref{fig_zdist_GISSEL} are identified as
early-type, in drastic contrast to Figure~\ref{fig_zdist}.  This
difference arises because the spectrum of an unreddened old stellar
population (i.e. an early type galaxy) at $z\approx 1.8$ looks very
similar, under the coarse resolution afforded by broadband filters, to
that of a reddened, Lyman-suppressed, star-forming galaxy at $z\approx
1.2$.  Again, as with the determination of redshift, we believe that
the extended CWW templates are more realistic and will produce more
accurate spectral types.

In the case of the pure GISSEL results the conclusion likely to be drawn is
that at $z\approx 2$ there exists a large population of old galaxies.  In
contrast, results based on extended CWW templates favor no such
radical conclusion.  Clearly, the true shape of galaxy SEDs in the UV
can have profound effects on the redshifts and spectral types that one
fits and, consequently, on the conclusions that are reached regarding
galaxy evolution.

\subsection{Our Best Template Set and the Redshifts it Produces}

We have limited ourselves to F814W$_{\rm AB} \leq 27.0$ which ensures
that aliasing due to photometric errors is insignificant.  We have
interpolated finely enough in SED, redshift, and normalization (the
$s$ of equation (1)) to ensure that template sparseness is not a
concern.  The third source of possible aliasing is the uncertainty in
the UV SEDs, especially of high redshift galaxies.  There is not much
that can be done here other than to say that by using empirical
template SEDs we are accounting for internal reddening, and that by
including the high-$z$ Lyman absorption we are accounting for
intergalactic hydrogen.  We also draw attention to the fact that we
recover the redshifts of the great majority of objects for which
spectroscopic redshifts exist (see Figure~\ref{fig_zphot_vs_ztrue}).
We are prepared to trust the validity of our results out to
$z\approx3.5$.  We have little confidence in any redshifts at $z>4$,
as there are no spectroscopic redshifts there with which we can verify
our photometric ones.

Armed with some degree of confidence in the photometric redshifts
obtained with the extended CWW templates, we now proceed to compute the
luminosity functions and luminosity densities of galaxies in the HDF.

\section{LUMINOSITY FUNCTIONS}\label{lumfuncs}

The luminosity function (LF) is a standard and basic way of describing 
the galaxy population.  
Using the redshift catalog obtained with the extended CWW template set
(Section \ref{technique}), we now turn to investigate the galaxy
population and its evolution by means of luminosity functions.  To
determine the LFs we use a maximum likelihood method (see, e.g.,
\cite{efs88} and \cite{lin97} for more details).  This method is
insensitive to fluctuations in galaxy density --- an important feature
in a pencil-beam survey such as the HDF.  To facilitate comparison
with $z<1$ results our LFs are computed in the rest-frame F450W$_{\rm
AB}\approx B_{\rm AB}$ band.  We fit the LF to the usual Schechter
(1976) parametrization and list the fit parameters ($M^*, \alpha$, and
$\phi^*$) in Table~\ref{tbl-1}.


Our sample has constant apparent magnitude limits and yet covers an
enormous baseline in redshift.  A consequence of the enormous redshift
range is the large K-corrections which effectively
remove red objects out of the sample at high redshift.  On the other
hand, blue, star-forming galaxies have much smaller K-corrections and
do not suffer from this effect.  It should be noted, however, that at
high enough redshifts one does not expect to see many intrinsically red
objects, since at those redshifts the universe may be too young for
stellar populations to have aged sufficiently.

\subsection{Comparison with $z<1$ Spectroscopic Surveys}

At $z<1$ we can compare our HDF 
luminosity functions with those obtained from spectroscopic redshift
surveys.  Figure~\ref{figlf1} shows such a comparison with the
$B_{\rm AB}$-band LFs derived from the Canadian Network for
Observational Cosmology (CNOC; \cite{lin97}; Lin et al., in
preparation) and Canada-France (CFRS; \cite{lil95} \footnote{Because
LFs for the particular comparisons we wanted to make were not
given in the CFRS LF paper (\cite{lil95}), we computed them ourselves
based on redshift catalogs and other data kindly provided by Simon
Lilly.})  redshift surveys.  The comparison is made against a sample
of CNOC galaxies with $0.12<z<0.6$ (1236 objects) and a sample of CFRS
galaxies with $0.5<z<1$ (424 objects).  The LFs shown in
Figure~\ref{figlf1} are those for all objects, irrespective of their
spectral types.  The agreement between the HDF LFs and those from the
CNOC and CFRS samples is good over the magnitude range common to the
HDF and the spectroscopic survey samples.  In this common range the LF
shapes from the three samples are similar, showing flat faint-end
slopes ($\alpha \approx -1$), though the normalization of the HDF $0.5
< z < 1.0$ LF is somewhat higher than that of the corresponding CFRS
sample. 

In the two upper panels of Figure~\ref{figlf2} we divide the HDF,
CNOC, and CFRS samples into subsamples of galaxies bluer or redder
than our CWW Scd model, and then compare the resulting LFs.  Our HDF
results are again consistent with those derived from the spectroscopic
surveys.
The agreement between the HDF
photometric redshift LFs and those derived from the spectroscopic
surveys is very encouraging, and gives us confidence to proceed
to fainter objects and to higher redshifts.

\subsection{Low-$z$ Faint Galaxies}


At absolute magnitudes fainter than those accessible to either CNOC or
CFRS, the HDF LF is no longer flat (see Figure~\ref{figlf1}), but has
a much steeper slope ($\alpha \approx -1.3$).  This excess of faint
galaxies is similar to that seen at $z\approx 0$ in the Center for
Astrophysics (CfA) survey (\cite{mar94a}; \cite{mar94b}) and in the
CFRS sample at $z < 0.2$ (\cite{lil95}), although the HDF data
demonstrate that the excess continues to at least as faint as $M_{{\rm
F450W}_{\rm AB}}=-11$.

When we split the LF by spectral type, as in the bottom panel of
Figure~\ref{figlf2}, we see that those faint galaxies which are
responsible for the steep $\alpha$ also have the blue late-type colors
indicative of ongoing star formation. The same trend was also reported
by Marzke et al.\ (1994a) for their local faint-end excess. About 80\%
of our sample at $0.2 < z < 1.0$ are actually bluer than our model
extended CWW Scd galaxy. 

In principle, the misidentification of HII regions as faint galaxies
(Section~\ref{data}) could cause an artificial
steepening of the faint-end slope.  However, the rate of misidentifications
(30\% assuming that \emph{all} non-isolated faint objects are
misclassified HII regions) is too small to negate the existence of the
star-forming dwarf excess.

The rather large redshift noise ($\sigma_z\approx 0.13$ at these
redshifts) could cause an artificial brightening of $M^*$ and
steepening of $\alpha$ (\cite{sub96}).  This is an effect akin to one
that photometry errors can have on the LF (e.g.,
\cite{efs88}).  By means of Monte Carlo simulations, we tested
the distortion that redshift noise may have on our determination of
the LF.  Specifically, we generate model redshift catalogs using some
given input LF, then perturb these catalogs with Gaussian redshift
errors similar to the noise we see in the actual HDF redshift catalog,
and finally compute the LF based on the perturbed model catalogs.  We
find that the effects of redshift errors are small (given the absolute
magnitude ranges we fit), with negligible effect on $\alpha$, and
slight brightening of $M^*$ (a couple tenths of a magnitude).
These effects are smaller than the 1$\sigma$ errors we quote in
Table~\ref{tbl-1}.  In particular, a flat ($\alpha = -1$) input LF
cannot steepen to the slope $\alpha=-1.3$ that we actually see in the
HDF.  We therefore conclude that the steep faint-end slope of our HDF
LF is not an artifact of photometric redshift errors.  We also
verified that the effects of redshift noise on the LF are not
important at higher $z$; for simplicity we will not make any
corrections for this effect in calculating our luminosity functions.

\subsection{Evolution of the Luminosity Function out to $z=4$}

We now turn to investigate the evolution of the luminosity function
with redshift.  The LFs for five redshift bins between $z=0.2$ and
$z=4$ are shown in Figure~\ref{figlf3}.  The LFs have not been split
by color but we note that the galaxies at higher redshifts are
almost exclusively blue, star-forming objects.  Schechter
function fits to the LFs are shown as lines and the fit parameters are
listed in Table~\ref{tbl-1}.  The $0.2<z<0.5$ (dashed line) and
$1<z<2$ (dotted line) fits are used as fiducials.

The LFs presented in Figure~\ref{figlf3} show clear signs of evolution
with redshift.  The bright end of the LF appears to brighten by $\sim
0.5$ mag from $0.2<z<0.5$ to $1<z<2$, and by a further $\sim 0.5$
mag from $1<z<2$ to $2<z<3$.  The faint end slope steepens
considerably with lookback time, from $\alpha = -1.3$ at $0.2<z<0.5$
to $\alpha = -2$ at $2 < z < 3$.  The HDF LF thus appears, almost
monotonically, to both brighten and steepen in shape with increasing
lookback time, up to $z\approx 3$.  The most drastic change
occurs between the $2<z<3$ and $3<z<4$ redshift bins, as there we see
the LF fade back to values similar to those seen at low redshift.  We
interpret this fading as consistent with the view that the majority of
present-day $\sim L^*$ galaxies began star formation around that time,
as we elaborate in Section
\ref{lumdens}.

We note that we do not see the $\sim
5$ magnitude brightening of the luminosity function measured by Gwyn
\& Hartwick (1996) and Mobasher et al.\ (1996).  The difference arises
because both internal reddening (absent from the models of Gwyn \&
Hartwick 1996) and Lyman absorption (absent from Gwyn \& Hartwick 1996
and from Mobasher et al.\ 1996) suppress the UV flux.  In the absence
of this suppression galaxies tend to be assigned much earlier spectral
types
\emph{even if} their redshifts are identified correctly.  Early
spectral types require very large K-corrections compared to those
needed for late types and consequently tend to be given much brighter
absolute magnitudes.  This effect manifests itself as an overly strong
brightening of the luminosity function.


\section{LUMINOSITY DENSITY}\label{lumdens}

\subsection{The Luminosity Density of the Universe}\label{lumdens1}

\begin{figure}
\epsscale{1.0}
\plotone{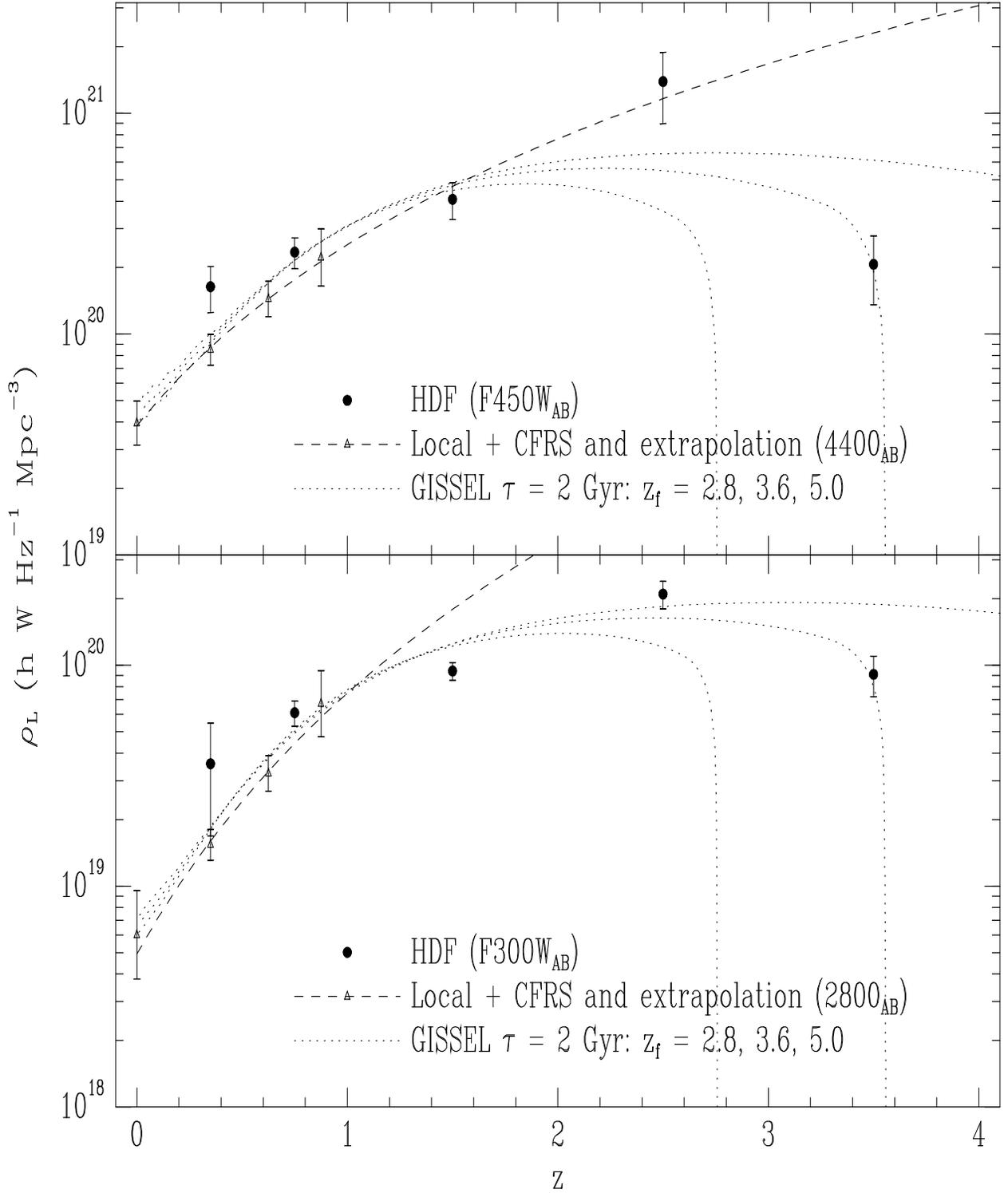}
\caption{
 Luminosity density of the universe from the HDF
 (filled symbols) at rest-4500\AA\ (top panel) and rest-3000\AA\
 (bottom panel) .  Open triangles are data from Lilly et al.\ (1996)
 and the dashed lines are fits (and extrapolations) of these data.
 Dotted lines are luminosity densities generated from GISSEL models
 assuming an exponentially decaying burst of star formation (see text
 for more details), which started at $z_f=$ 2.8, 3.6, or 5.0.}
\label{figlumdens}
\end{figure}

We obtain the comoving luminosity density as a by-product of the
luminosity function measurements.  Figure~\ref{figlumdens} shows the
comoving luminosity densities measured at rest-frame F450W$_{\rm AB}$
and F300W$_{\rm AB}$.  The luminosity density in each redshift bin was
computed by integrating the corresponding LFs over $-23 <M_{\rm
F450W_{AB}} < -15$ or $-21 < M_{\rm F300W_{AB}} < -15$; the errors
were estimated from the standard deviation of the mean of the three
Wide Field Camera chips.  The values for the $z<1$ luminosity
densities obtained by Lilly et al.\ (1996) from the CFRS at similar
wavelengths (4400\AA , 2800\AA) are also shown, together with their
$(1+z)^{\beta}$ parametrizations\footnote{Lilly et al.\ (1996) give
$\beta=2.7$ at 4400\AA\ and $\beta=3.9$ at 2800\AA\ in a $q_0=0.5,
\Omega_0=1$ universe}.  
Up to $z\approx 3$, the HDF luminosity densities show a monotonic
increase in both the rest- F300W and F450W bands, although (at least
for the F300W band) not at as high a rate as one would extrapolate
from the the CFRS $z<1$ data.
The luminosity density peaks in the $2 < z < 3$ bin and past $z=3$ it drops.
These same trends were also seen in the LF
evolution illustrated in Figure~\ref{figlf3}.  

It is interesting to interpret the evolving luminosity density in
terms of the global star formation history of the universe.  We use
simple GISSEL models to represent the star formation history of all 
galaxies within a comoving volume element.
In Figure~\ref{figlumdens} we choose one particular set of models
synthesized from the GISSEL library to compare against the observed
luminosity densities.  The models are those for a stellar population
characterized by a Salpeter initial mass function (IMF) ($x=1.35$,
$0.1 M_{\odot}<M<125 M_{\odot}$) in which star formation exponentially
declines with a decay time $\tau=2$~Gyr (\cite{bru93}).  As in the
rest of this paper, we use $q_0 = 0.5$, but here we choose a specific
$H_0$ value, 45~km~s$^{-1}$~Mpc$^{-1}$, in order to make $t_0 =
14.5$~Gyr.
We choose to
start the star formation in our models at $t_f = 1, 1.5$, and 2 Gyr
after the Big Bang, corresponding to formation redshifts $z_f = 5.0,
3.6$, and 2.8, respectively.  The models were arbitrarily normalized
by eye to pass between the low-$z$ HDF points (at $0.2 < z < 1.0$) and
the $z < 1$ CFRS results.  

All three models are quite similar at $z
\lesssim 1.5$ and reproduce the observed luminosity density trends
well at $z < 1$. At high redshifts the $z_f = 3.6$ model appears to 
match the observations best.  It is intriguing that this simple toy
model reproduces, albeit roughly, the bulk features of the
observed luminosity density evolution in the HDF, and that it does so
in both bandpasses at the same time.  In the context of these toy
models, the implication of our results is that the bulk of the
stellar population of the universe appears to have started forming
at $z = 3-4$.  Clearly, our models are almost
certainly an oversimplification of reality, but they do give us a
rough handle on the initial epoch of galaxy formation.

\subsection{Production of Metals from $z=4$ to the Present}

The UV flux from galaxies can be used as a measure of both the star
formation rate and the metal production rate (e.g., \cite{cow88};
\cite{son90}; \cite{cow96}).  The metal production rate 
can be more readily related to the UV flux, with the results
independent of cosmology, and relatively independent of the details of
galaxy or star formation history or the specific form of the IMF
(\cite{cow88}).  As an interesting consistency check on the results
presented earlier, we investigate the metal density of the universe
(inferred from the HDF UV luminosity densities) as a function of
redshift.

Songaila et al.\ (1990) give the relevant relation 
\begin{equation}
\frac{\rho Z \ } {10^{-34} \ {\rm g \ cm}^{-3}} = 
  \frac{S_\nu \ }{ 3.6 \times 10^{-25} \ {\rm ergs \ cm}^{-2} \ {\rm s}^{-1}
  \ {\rm Hz}^{-1} \ {\rm deg}^{-2}} \ ,
\end{equation}
where $\rho Z$ is the present volume density of metals produced by
some population of objects at redshift $z$.  $S_\nu$ is the present
observed surface brightness of those objects at observed frequencies
$\nu = \nu_0 / (1+z)$. The source frequencies $\nu_0$ have to lie in a
flat (in $f_\nu$) part of the source spectrum, which is the case in
rest-UV for star-forming galaxies. We can thus transform 
rest-${\rm F300W}_{\rm AB}$ luminosity densities to presently-observed
surface brightnesses, and then infer the present-day density of metals
that were produced by these high-$z$ star-forming HDF galaxies.

In Figure~\ref{figmetal} we plot (solid line) the cumulative
metal density produced by HDF galaxies as a function of
$z$.  Here we have assumed that the universe was metal-free before $z = 4$
(as we do not trust our photometric redshifts beyond $z = 4$, we have
no constraints there).
The bounds on the present-day metal density of the universe, 
$\rho Z = 8 - 24 \ h^2 $~g~cm$^{-3}$ (\cite{cow96}), are shown as dashed lines
for the Hubble constant $h = 0.7$.  
The lower and upper bounds for $h = 0.5$ and $h =
1$, respectively, are shown as dotted lines.  
By $z = 0$ the inferred metal density matches the
local constraints well, implying that we are indeed seeing in the HDF
all the needed metal production. Also shown (open squares) are the
metallicity constraints from high-redshift observations of damped
Lyman-$\alpha$ systems made by Pettini et al.\ (1994) and Lu et al.\
(1996); the large error bars on the data are meant to represent the 
large spread in metallicity values that is observed.  To convert
the metallicity measurements to actual metal densities we assumed that
the mass density in damped Lyman-$\alpha$ systems at high redshifts is
the same as that in present-day stars (\cite{lan95}); that is, that
the damped Lyman-$\alpha$ systems are indeed the progenitors of
present day galaxies. (Here we also used the central value $h = 0.7$.)
Keeping the large uncertainties in mind, there is 
reasonable agreement between the metal densities inferred from the
HDF and from damped Lyman-$\alpha$ systems.  It is reassuring that the
redshift-dependent metal density derived from the F300W$_{\rm AB}$
luminosity density, which in turn was derived using
\emph{photometric} redshifts, is consistent with both the local and 
high-redshift metal densities obtained from completely independent 
measurements.

\section{DISCUSSION}\label{discussion}

As discussed in Section~\ref{lumdens1}, our simple GISSEL toy model
provides a rough but reasonable match to the observed evolution of the
HDF luminosity density. Within the context of this model, the rise in the 
luminosity density 
from $z\approx 3.5$ to
$z\approx 2.5$ means that the initial major epoch of star formation in galaxies
occured at $3<z<4$. Such a picture is certainly consistent with the presence
of a population of luminous, star-forming galaxies discovered recently
at $z \sim 3$ (e.g., \cite{ste96}; \cite{yee96}).  
Also, interestingly, the
peak in the luminosity density seen in the $z = 2$--$3$ bin 
coincides in redshift with the peaks observed in the
number density of bright, optically-selected quasars (e.g.\ \cite{war94})
and radio galaxies (e.g. \cite{dun90}).

\begin{figure}
\epsscale{0.5}
\plotone{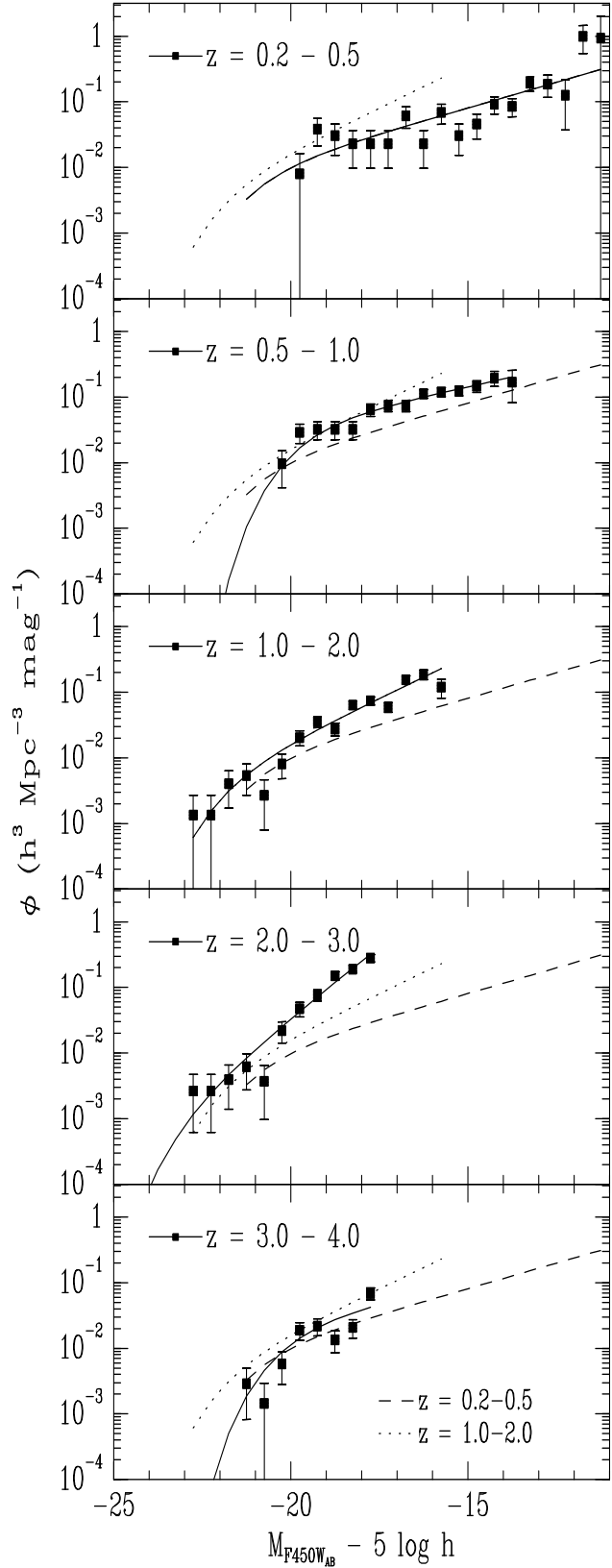}
\caption{
Evolution of the luminosity function with redshift.
 Solid lines are Schechter function fits to the data.  Dashed and
 dotted lines are fiducial LFs --- the dotted line is the $1<z<2$ HDF
 LF and the dashed line is the $0.2<z<0.5$ HDF LF.  Schechter fit
 parameters are listed in Table~\ref{tbl-1}.}
\label{figlf3} 
\end{figure}

Below $z \approx 3$ we see several conspicuous trends in the HDF
luminosity density and luminosity function. There is a strong decline
in the luminosity density with decreasing redshift. This decline is a
reflection of the accompanying marked changes in the LF: the LF is
simultaneously fading and flattening at ever lower redshifts, as seen
in Figure~\ref{figlf3}.  The flattening of the LF faint-end slope is a
characteristic of hierarchichal models of galaxy formation
(e.g. \cite{col94}).  In our data this flattening appears to be
accompanied by moderate fading of the bright end of the LF, perhaps
indicating that both merging and luminosity evolution play a role.

As the majority of galaxies we
observe in the HDF are star-forming objects, the changes seen
in the LF imply a migration in the characteristic luminosities of
star-forming galaxies, from brighter at high $z$, to fainter at lower 
redshifts. 
If we can roughly associate the $B$-band luminosities of
these galaxies with their underlying stellar masses (though it would have been
preferrable to have longer rest-wavelength data, such as $K$-band, for
this purpose;
\cite{cow96}), our LF results suggest that the less massive the
galaxy, the more recently it is undergoing a period of strong star
formation.  This same trend has been observed (and termed
``downsizing'') in the spectroscopic survey sample of Cowie et al.\
(1996) over lower redshifts ($0.2 < z < 1.7$).  Our analysis
indicates that this trend extends over the entire redshift range from
$z \sim 3$ to the present; the characteristic star-forming galaxy
luminosity migrates from the bright end of the LF at $z \sim 3$ to
below present-day $M^*$ at $z < 1$.  At $z<0.5$, star-forming galaxies
dominate the dwarf population.  If star-forming galaxies are indeed in
the process of initially assembling themselves, then we may
conclude that galaxy formation occured sequentially in size, with the
largest objects forming at $z>3$ and the smallest only recently.  This
picture is consistent with the scenario of galaxy formation reviewed
by Fukugita et al.\ (1996), in which massive spheroidal galaxies
formed at $z\approx 3$, followed at later epochs by the formation of
less massive objects, and finally by the formation of dwarfs in recent
times.  That the mass density of neutral gas in damped Lyman-$\alpha$
systems (the likely progenitors of present-day galaxies) appears to
decrease rapidly below $z \approx 3.5$ (\cite{lan95}) lends support to
the picture that conversion of gas to stars was actively happening
below $z \sim 3$. Moreover, as a consistency check on our results, the
density of metals inferred from the HDF agrees with the
bounds available at both low and high redshifts
(Figure~\ref{figmetal}).

\begin{figure}
\plotone{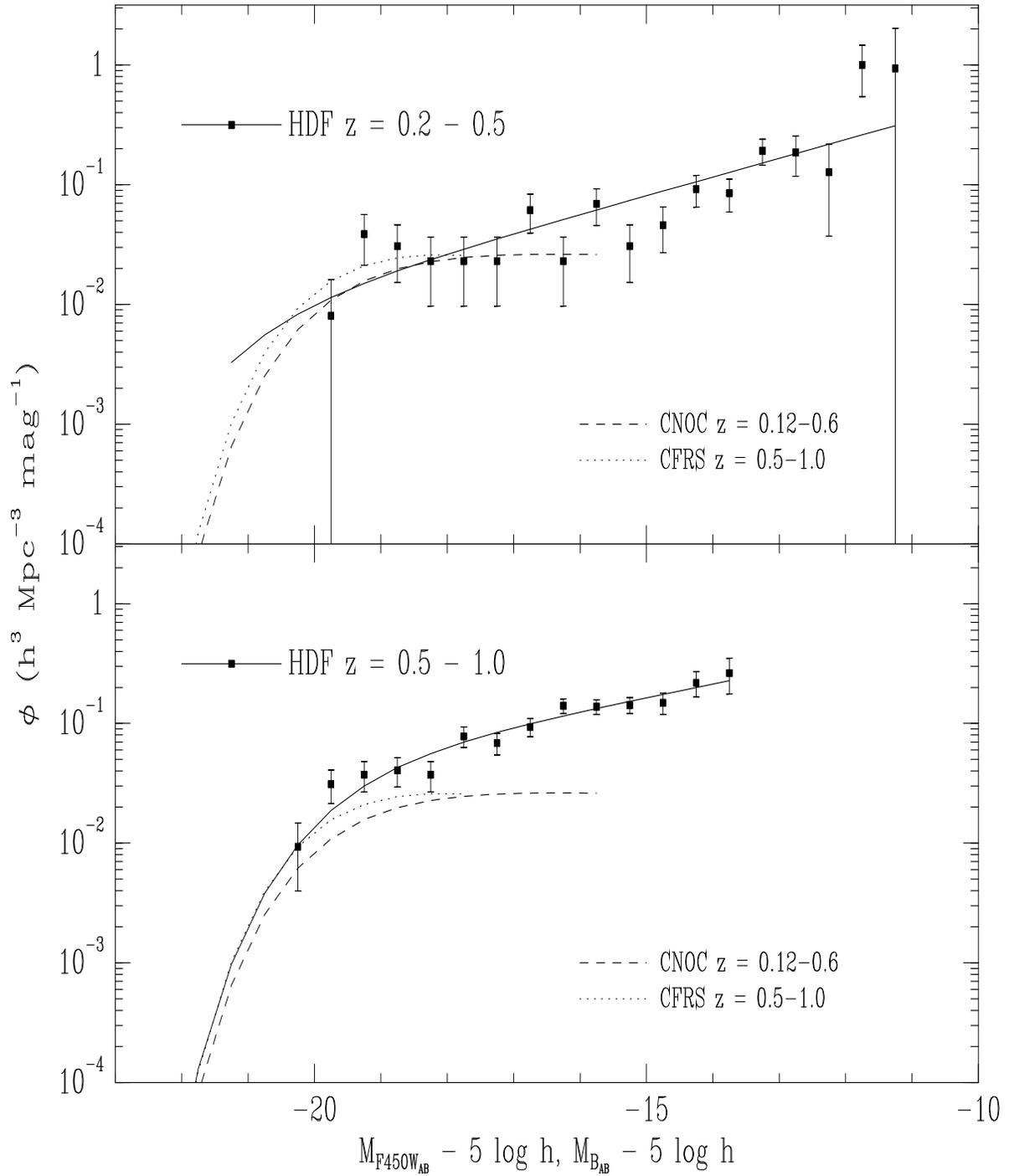}
\caption{
Comparison of HDF luminosity functions with those
 from the CNOC and CFRS spectroscopic redshift surveys.  Top panel:
 $0.2<z<0.5$.  Bottom panel: $0.5<z<1.0$.  Solid lines are Schechter
 function fits to the HDF data; fit parameters are listed in
 Table\ref{tbl-1}.}
\label{figlf1} 
\end{figure}

\begin{figure}
\plotone{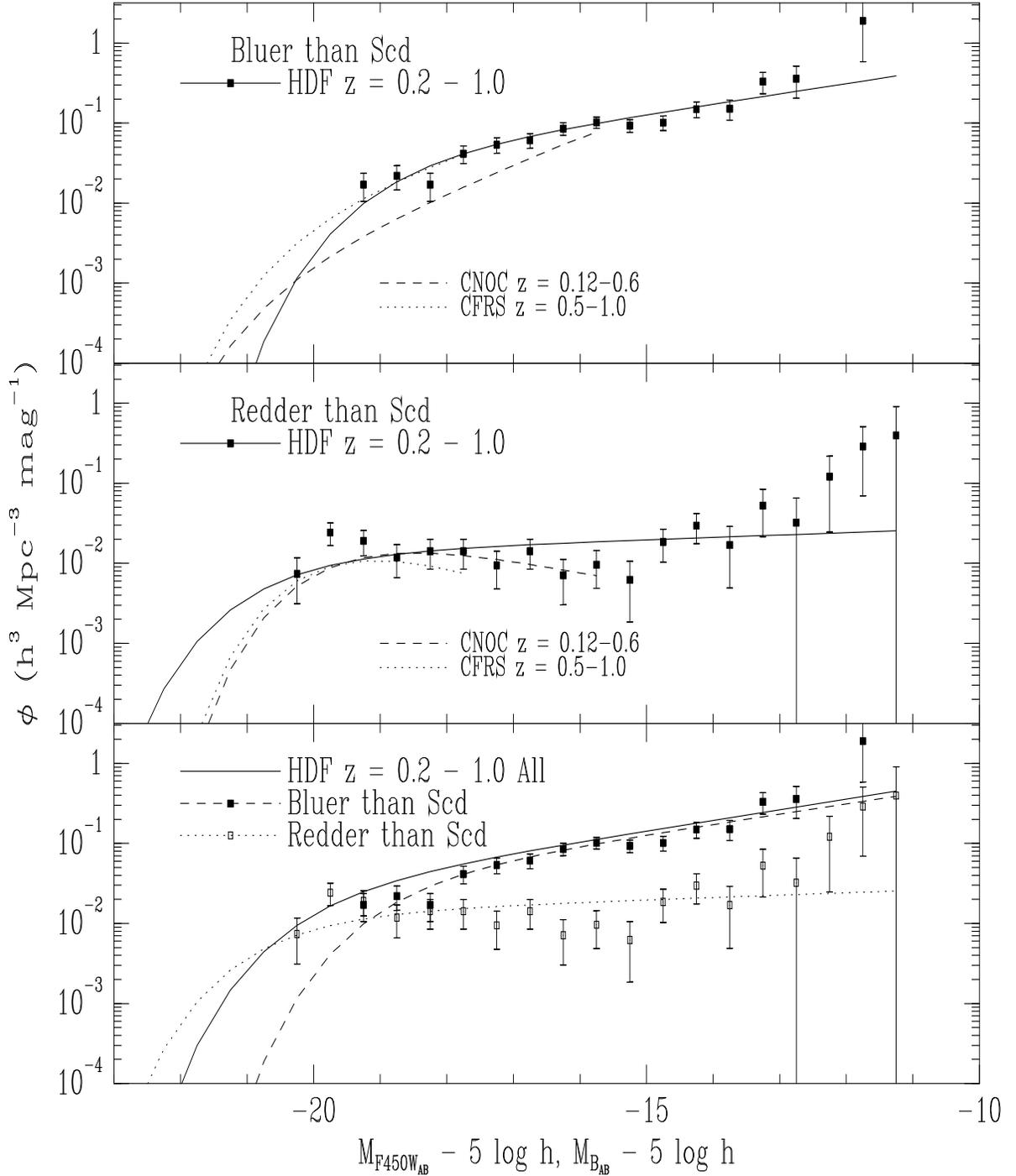}
\caption{
HDF luminosity functions split by color.  The top
 two panels compare the HDF LFs with CNOC and CFRS spectroscopic
 survey results when split into blue (top panel) and red (middle
 panel) subsamples.  The bottom panel compares the relative
 contributions of galaxies bluer than Scd (filled symbols) and those
 redder than Scd (open symbols); the solid line is the Schechter
 function fit to all the $0.2<z<1.0$ HDF objects.  }
\label{figlf2}
\end{figure}

At the lowest redshifts, $z<1$, our luminosity function and luminosity
density results agree, within regions of overlap, with results from
spectroscopic surveys (Figures~\ref{figlf1} and \ref{figlf2}).
Furthermore, as an extension of the downsizing trend from higher $z$,
we see a large population of star-forming dwarfs extending to
luminosities below the reach of the current spectroscopic surveys
(Figure~\ref{figlf2}).  Such low-redshift bursting
dwarfs have been evoked (e.g., \cite{bro88}) to explain the excess
counts of faint blue galaxies.
Recent evidence
from direct luminosity function measurements (\cite{ell96};
\cite{cow96}; \cite{lil95}), as well as more indirect morphological
studies of deep HST images, including the HDF (e.g., \cite{gla95a};
\cite{abr96}), have indicated that the faint blue galaxies are rapidly
evolving, star-forming, and have irreguliar or peculiar morphologies. 
Our direct HDF luminosity function results lend further support to this picture, 
and also show that the low-$z$ dwarf population reaches very faint
luminosities indeed.

Although we have outlined above an overall picture of the
star-formation history of the universe, many of the specifics remain
to be filled in.  Are the LF changes, in particular the steepening of
$\alpha$ at higher $z$, due to luminosity-dependent luminosity
evolution, or is merging also taking place?  What are the individual
evolutionary tracks of particular galaxies in the space of luminosity,
spectral type, morphology, and other parameters?  Will detailed galaxy
formation models, which include both galaxy, stellar, and clustering
evolution (e.g., \cite{kau95}; \cite{bau96}), be able to account for
the clear redshift-dependent changes we observe in the HDF luminosity
functions?  These and similar questions have not been addressed in the
present paper, though we do plan for future work to relate the
morphologies of HDF galaxies to the existing LF information, thereby
hopefully filling in more of the details in the picture of galaxy
evolution.

We caution that our results rely on the validity of photometric
redshifts --- redshifts whose determination, as has been shown in
Section \ref{aliasing}, can be fraught with danger.  The fact that our
redshift determination relies on the poorly constrained UV properties
and equally poorly constrained spectral evolution of galaxies is a
major cause for concern.  Although we have taken many precautions to
guard against catastrophic redshift errors, our results and the
conclusions that we derive from them can be confirmed only with
additional spectroscopic redshift verification.


\vspace{1cm}
We thank Bob Williams for assigning Director's Discretionary Time to
the Hubble Deep Field and the HDF team for the data.  We also thank
David Schade and Gabriela Mall\'en-Ornelas for useful comments and
Nick Kaiser for the encoragement he provided.  This research was
supported by NSERC of Canada.

\clearpage

%
%

\clearpage

\begin{deluxetable}{cccccr}
\tablewidth{0pt}
\tablenum{1}
\footnotesize
\tablecaption{Parameters of Schechter Function Fits \label{tbl-1}}
\tablehead{
\colhead{$z$ range}   & \colhead{color sample}   & 
\colhead{$M^*_{\rm F450W_{AB}} - 5 \log h$}   & 
\colhead{$\alpha$}   & \colhead{$\phi^*$ ($h^3$~Mpc$^{-3}$)} &
\colhead{$N$} 
}
\startdata
0.2 -- 1.0  & all & $-20.1 \pm 0.3$ & $-1.33 \pm 0.05$ & $0.033 \pm 0.009$
& 357 \nl
0.2 -- 1.0  & bluer than Scd & $-18.9 \pm 0.3$ & $-1.3 \pm 0.1$ & $0.045 \pm 0.012$
& 277 \nl
0.2 -- 1.0  & redder than Scd & $-20.7 \pm 0.6$ & $-1.1 \pm 0.1$ & $0.015 \pm 0.006$
& 80 \nl
\nl
0.2 -- 0.5  & all & $-21.2 \pm 1.6$ & $-1.4 \pm 0.1$ & $0.009 \pm 0.007$
& 103 \nl
0.5 -- 1.0  & all & $-19.9 \pm 0.4$ & $-1.3 \pm 0.1$ & $0.042 \pm 0.013$
& 254 \nl
1.0 -- 2.0  & all & $-22.1 \pm 0.7$ & $-1.6 \pm 0.1$ & $0.006 \pm 0.006$
& 291 \nl 
2.0 -- 3.0  & all & $-23.2 \pm 1.5$ & $-2.1 \pm 0.1$ & $0.002 \pm 0.004$
& 198 \nl 
3.0 -- 4.0  & all & $-20.4 \pm 0.7$ & $-1.3 \pm 0.3$ & $0.023 \pm 0.017$
& 63 \enddata
\end{deluxetable}


\clearpage

\begin{figure}
\includegraphics{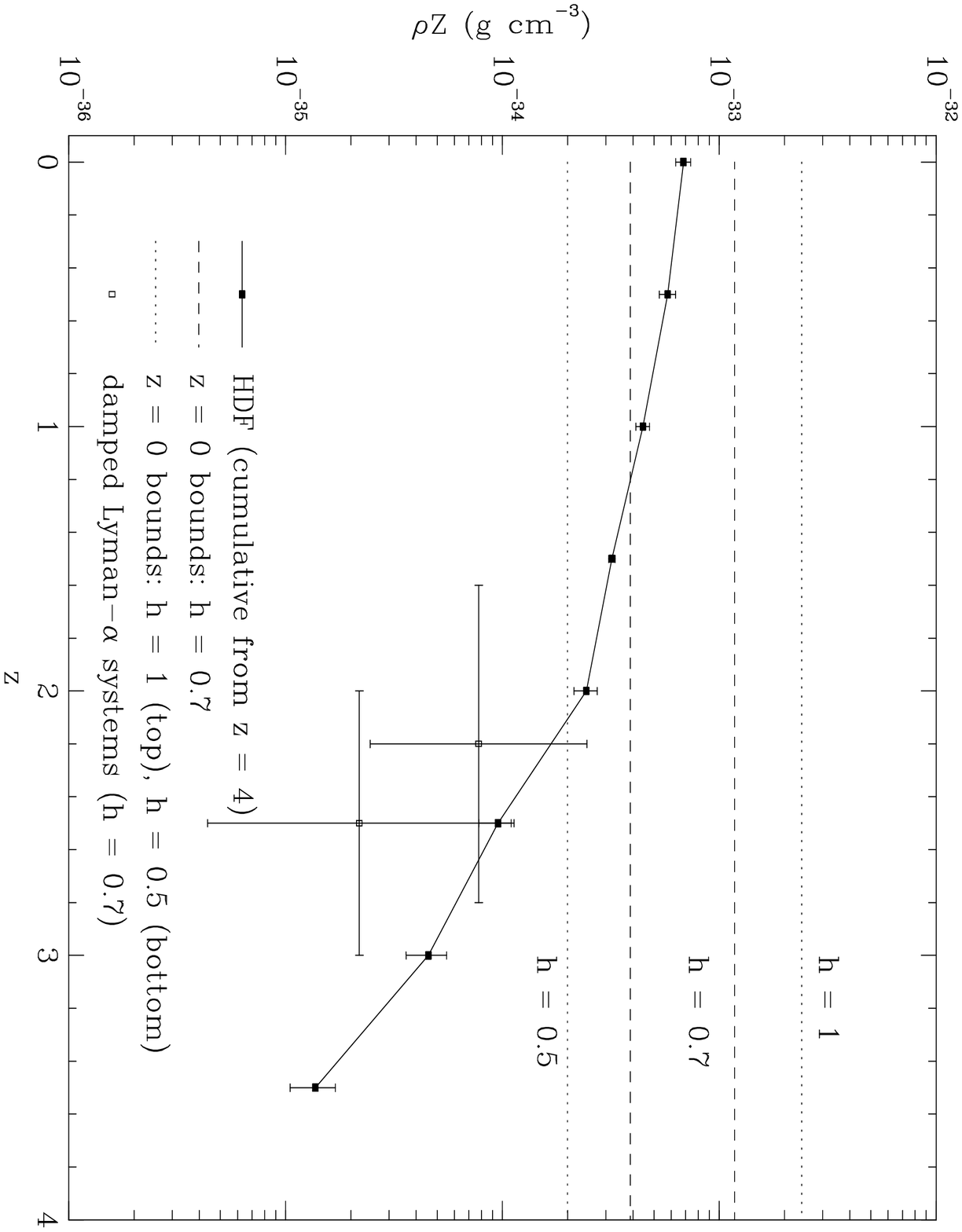}
\end{figure}
\vspace{30cm}
\begin{figure}
\caption{
The metal density of the universe derived from UV
 fluxes of HDF objects (filled squares).  Open squares are metal
 densities inferred from metallicities of damped Lyman-$\alpha$
 systems.  Dashed lines represent bounds on the present-day metal
 density assuming $h=0.7$; dotted lines are upper and lower bounds
 assuming $h=1$ and $h=0.5$ respectively (see text for more
 details).}
\label{figmetal} 
\end{figure}

\end{document}